# Large magnon dichroism and other optical properties of hexagonal ferrite $h$-Lu$_{0.6}$Sc$_{0.4}$FeO$_3$ with altermagnetic A$_2$ spin ordering

V. A. Martinez and A. A. Sirenko
*Department of Physics, New Jersey Institute of Technology, Newark, New Jersey 07102, USA*

L. Bugnon, P. Marsik, and C. Bernhard
*Department of Physics, University of Fribourg, CH-1700 Fribourg, Switzerland*

Qing Zhang
*School of Physics, Shandong University, Jinan 250100, China*

G. L. Pascut
*MANSiD Research Center and Faculty of Forestry, Stefan Cel Mare University (USV), Suceava 720229, Romania*

F. Lyzwa
*Department of Physics, New Jersey Institute of Technology, Newark, New Jersey 07102, USA and NSLS-II, Brookhaven National Laboratory, Upton NY, USA*

Z. Liu
*Department of Physics, University of Illinois at Chicago, Chicago, Illinois 60607-7059, USA*

K. Du and S.-W. Cheong
*Keck Center for Quantum Magnetism and Department of Physics and Astronomy, Rutgers University, Piscataway, New Jersey 08854, USA*

Multiferroic hexagonal $h$-Lu$_{0.6}$Sc$_{0.4}$FeO$_3$ single crystals with non-collinear spins were studied using the THz and Raman scattering spectroscopies and ellipsometry. Antiferromagnetic resonances, or magnons, were found at about 0.85 THz and 1.2 THz. These magnons harden as temperature increases and disappear above 130 K. This behavior is consistent with the magnetic susceptibility and a phase transition to a previously reported weak ferromagnetic state. A strong dichroism at the resonance with the AFM doublet has been observed at zero external magnetic field using both conventional circular polarization and THz vector vortex beams. This observation is attributed to the strong altermagnetic properties of $h$-Lu$_{0.6}$Sc$_{0.4}$FeO$_3$ with a broken $\boldsymbol{P} \cdot \boldsymbol{T}$ symmetry. The splitting of the magnon doublet in an external magnetic field applied long the $c$ axis yields a $g$-factor of 3.0 for the Fe$^{3+}$ ions. Raman spectra of the optical phonons revealed a Fano-type asymmetry due to their interaction with a continuum of polar excitations. Electronic transitions were studied with ellipsometry and the results were compared with the modelled using DFT+eDMFT.

(*) Author to whom correspondence should be addressed: vladimir.martinez@njit.edu

**INTRODUCTION**

Multiferroic hexagonal ferrites $h$-(Lu,Sc)FeO$_3$ (LSFO) are improper ferroelectrics that are structurally similar to the better-known family of hexagonal rare earth ($RE$) manganites $h$-$RE$MnO$_3$ and hexoferrites $h$-$RE$FeO$_3$. The primary distortion in these parent materials is a single unstable mode with $q$=(1/3 1/3 0) (the $K$ mode) consisting of a rotation of the FeO$_5$ or MnO$_5$ trigonal-bipyramids and a buckling of the $RE$ planes, resulting in a tripling of the unit cell volume.[1] When the $K$ mode condenses at the structural transition temperature $T_C$, which is usually above 1000 K, the ratio of the $RE$ ions with upward distortions and those with downward distortions become 2:1. This causes a ferroelectric polarization along the $c$ axis through non-linear coupling, simultaneously the Fe$^{3+}$ ions form trimers in the $a$-$b$ planes. The ferroelectric displacements are mostly driven by the ionic size effect, and therefore these and related materials are also referred to as geometric ferroelectrics.[2] The resulting broken-inversion-symmetry crystal structure has the $P6_3cm$ space group.[3,4] The electronic origin of ferroelectricity in LSFO was studied in Ref.[5] using soft X-ray spectroscopy.

The Fe ions form a trimerized triangular lattice and the Fe spins order in the 120° structure below $T_N$(Fe)=160 K. In $h$-$RE$FeO$_3$ three trimerization antiphase domains are interlocked with two ferroelectric up-down domains in such a way that all possible six structural domains merge to one point forming topological vortex-antivortex domains.[6,7] While the typical magnetic state of the better known manganite analogues $h$-RMnO$_3$ is the so-called B$_2$ state, $h$-LuFeO$_3$ exhibits a very different A$_2$ state (see **Fig. 1**). Because of the broken structural inversion symmetry, the A$_2$ state of $h$-LuFeO$_3$ acquires a small net magnetic moment along the $c$ axis for which the ferroelectricity and weak magnetism are coupled.[8,9]

When $h$-LuFeO$_3$ is doped either with Sc or Mn, the material becomes more stable, resulting in the growth of large single crystals that are convenient for systematic studies of the magnetic and lattice excitations. In theory, the LSFO system allows for competing magnetic orders allowing for both A$_1$ and A$_2$ phases to be realized,[10] which are sketched in **Fig. 1**. The interesting spin structures of such samples have already been studied with neutrons[1,11] and magnetic domain spectroscopies.[3,12] The interpretation of the experimental data is not straightforward because both spin structures can coexist forming micron-size domains, which are much smaller than the mm-sized foot-print of a THz beam. The temperature dependence of the magnetic susceptibility of LSFO reported in Ref. [3] indicates that the slightly ferrimagnetic A$_2$ structure is formed below $T_N$=160 K. Upon cooling below 60 K, the fraction of the non-ferromagnetic A$_1$ domains increases and the A$_1$ and A$_2$ phases thus coexist at low temperatures. Our main interest was to study the connection between the spin structures, A$_1$ and A$_2$, and the possible altermagnetic properties of LSFO. The A$_1$ phase is antiferromagnetic (AFM) with co-planar spins for which the $\boldsymbol{P \cdot T}$ symmetry is not broken.[13] Here $\boldsymbol{P}$ is parity, $\boldsymbol{T}$ is time reversal, and $\boldsymbol{P \cdot T}$ is parity times time reversal. Thus, the A$_1$ phase is not altermagnetic. The magnetic point group is $6/m'mm$ which has a magnetic toroidal moment. This phase is expected to have nonreciprocal light propagation for the $k$-vector of light in the $ab$-plane even when the light is unpolarized. In contrast, the A$_2$ spin structure has a broken $\boldsymbol{P \cdot T}$ symmetry due to co-directional canting ferromagnetic moment and electric polarization that are both directed along the $c$-axis. Such a combined spin and polar order can be classified as a "strong altermagnetic" phase. This classification holds in spite of the circumstance that the ferromagnetic moment in LSFO is driven by the spin-orbit coupling.[14,15,16]

Far-infrared, THz, and Raman spectroscopies of magnons can reveal details of the spin interactions and the local magnetic structure of materials. In this work we report on systematic studies of the AFM resonances, or magnons, using the THz and Raman scattering spectroscopies in hexagonal LSFO. In the literature Raman data for antiferromagnetic resonances are available only for orthorhombic LuFeO$_3$ where they indicate two orthogonally polarized modes with energies of 0.6 and 0.7 THz (18.5 cm$^{-1}$ and 22 cm$^{-1}$).[17] We demonstrate that the magnon spectra can be used as a probe of the magnetic order. In particular, we show that the selection rules for the preferred circular polarization of the magnons reveals nonreciprocal effects that provide evidence for the A$_2$ phase with simultaneously broken $\boldsymbol{P}$ and $\boldsymbol{T}$ symmetries and thus a dichroic signal even at zero external magnetic field (in contrast to the A$_1$ phase, where such dichroism is not possible). In addition, we studied the magnons of LSFO with topologically similar light probes, such as THz vector vortex beams (VVB). Complementary optical spectroscopy studies of the phonons and the electronic transitions have also been performed to aid in the interpretation and analysis of the magnon spectra.

## EXPERIMENT

The floating-zone growth technique was utilized to produce bulk crystals of hexagonal $Lu_{0.6}Sc_{0.4}FeO_3$. Single crystals were cut and mechanically polished in the form of platelets. The samples were oriented so that the $c$-axis of LSFO was either perpendicular or parallel to the sample surface. The in-plane cross-sectional areas were larger than the beam cross section of about $7 \times 7$ mm$^2$, and the 0.95-mm thickness of the samples along the $c$-axis was optimized to provide a sufficient absorption in the transmission measurements of the magnons. To minimize the interference fringes in the optical experiments, the opposite sides of the samples were wedged at an angle of about 0.2◦ by mechanical polishing.

The transmission experiments were carried out at the University of Fribourg using a time-domain THz spectrometer with a photoconductive antenna emitter and a ZnTe birefringent detector. The available spectral range was between 0.1 THz and 3.0 THz with a spectral resolution of 0.03 THz. A slow focusing of the THz radiation on the sample with the $f$- number equal to 10 was implemented with mirror optics. The raw transmission data through the sample were normalized to the transmittance through an empty aperture of equal size. For samples with strong thickness interference fringes we additionally normalized the transmitted intensity in the magnon spectral range to that measured at high temperatures or high magnetic fields, where the magnons are absent. The sample was mounted on a cold finger inside a JANIS superconducting optical magnet that enables variable magnetic fields between $B= -7$ T and $+7$ T applied in the Faraday configuration so that the directions of the light propagation and the field coincide.

The THz setup was equipped with broadband free-standing wire-grid linear polarizers and optical retarders made of undoped Si prisms. The retarders allow for the conversion from linear $(\vec{e}_x \pm \vec{e}_y)$ to the right- or left-hand circular polarizations $\vec{e}_R = \vec{e}_x - i \cdot \vec{e}_y$ and $\vec{e}_L = \vec{e}_x + i \cdot \vec{e}_y$. An axicon retarder made from undoped Si (by TYDEX) was used to produce broad-band vector vortex beams with the orbital angular momentum (OAM) $l= +1$ or $l= -1$ and the spin angular momentum σ=0 [**Fig.1(c)**]. The electric field distribution of the vector beam is calculated to be $\vec{e}_l(\vec{r},\phi) \approx (\vec{r}/r) \cdot \exp[i \cdot l \cdot (\phi - \phi_0)]$, where $\phi$ is the vortex beam phase, the initial phase is $\phi_0 = 3\pi/4$, and $\vec{r}$ is the radial coordinate. Decomposition of the Note here that in the very center of the axicon ($r$=0), the intensity and polarization are not well defined and it is not clear if the intensity is zero as expected for conventional Laguer-Gaussian vortex beams. To avoid this uncertainty in our experiments, the center of the axicon was masked, so no rays with $r$<2 mm were transmitted. In the long-wavelength limit with $r$ comparable to λ, the integral of the electric field over the axicon cross section is equivalent to circular polarization. But in the spectral range around 1 THz used here λ=0.3 mm while the axicon acceptance window has a diameter of 20 mm, so that most of the electric field distribution forms a VVB, as expected. More details about using axicons for conversion between conventional circular polarization to the vortex beams $\sigma\pm \rightarrow l\pm$ can be found in Ref. [18]. A Woollam VASE ellipsometer at the Center for Functional Nanomaterials of Brookhaven National Laboratory was used for measurements of the electronic transitions at room temperature. Spectra of the infrared-active optical phonons were measured using the in house far-IR ellipsometer at the University of Fribourg.[19] Raman scattering experiments were carried out at the National Synchrotron Light Source, Brookhaven National Laboratory at the 22-ID-FIS beamline equipped with a single grating spectrometer, 532 nm laser and a set of linear and circular polarizers. For the low-temperature Raman scattering measurements the LSFO samples were mounted inside an Oxford LHe flow optical cryostat.

## EXPERIMENTAL RESULTS

### A. Magnetic excitations studied with THz spectroscopy

Several weak magnetic modes have been found in the THz part of the optical spectrum. Their preferable polarization and strong temperature dependence allowed us to attribute two of them to magnons (M1 and M2) and another one to an electromagnon (EM). **Figure 2(a)** shows a transmission spectrum at $T$=6 K and zero magnetic field (red symbols) and a fit with two absorbing magnon modes (black line) at 0.8 THz (26.5 cm$^{-1}$) and 1.2 THz (39 cm$^{-1}$). The light propagation is along the $c$-axis and the sample thickness was 0.95 mm. A simple harmonic oscillator model was used with the corresponding oscillator parameters listed in **Table I**. These modes are attributed to magnons, or AFM resonances, because they are excited with the magnetic field component of the light $\vec{h}$. For the magnon-dominated absorption we considered the corresponding adjusted oscillator strength $S_T$ for each mode that consists of contributions from electric and magnetic dipoles as follows: $S_T = \left( \mu_\infty \cdot S_e + \varepsilon_\infty \cdot S_m \right)$.[20] Here $\mu_\infty = 1$ and $\varepsilon_\infty = 18$ are the corresponding parameters for LSFO in the frequency range above the magnetic modes, $S_e = 0$ for the pure magnetic modes, and $S_m$ is the magnetic oscillator strength $S_m = S_T / \varepsilon_\infty$, which is also listed in **Table I**. Note that the magnetic oscillator strength for the modes polarized in the $a$-$b$ plane is very weak relative to typical values of magnon modes in other AFM compounds [20]. That is why for detection of the magnetic modes we measured relatively thick samples with $d$ = 0.95 mm.

**Figure 2(b)** shows transmission spectra and results of the fit with two absorptions modes at 22.5 cm$^{-1}$ and 28 cm$^{-1}$. Linearly polarized light was used with $\vec{e} \parallel c, \vec{h} \parallel a$. The light propagation is along the $b$-axis and the sample thickness was about 1 mm. The mode at 28 cm$^{-1}$ is attributed to the same M1 magnon as in **Fig. 2(a)**, which is excited by $\vec{h}$ in the $a$-$b$ plane. Note that the linearly polarized light couples to both components of the magnon doublet that results in the frequency shift compared to the single mode shown in **Fig. 2(a)**. The mode at 22.5 cm$^{-1}$ is excited by the electric component of light $\vec{e} \parallel c$, but it has a magnetic origin based on its strong temperature dependence and disappearance at higher $T$ approaching $T_N$. Thus, we attribute the mode to an electromagnon (EM). The inset in **Fig. 2(b)** shows the temperature dependence of the EM frequency, which demonstrates its hardening upon cooling. A simple harmonic oscillator model was used to fit the experimental data with the oscillator parameters listed in **Table I**. The oscillator strength $S_T$ of the modes is determined in the units of $\varepsilon(0)$. The electric oscillator strength of the electromagnon at 22.5 cm$^{-1}$ does not require renormalization since $S_e = S_T$ for electric dipoles.

**Figure 2(c)** shows transmission spectra in the complementary geometry using linearly polarized light with $\vec{e} \parallel a, \vec{h} \parallel c$. The light propagation is along the $b$-axis, the same as in **Fig. 2(b).** The absence of the M1 mode at 26.5 cm$^{-1}$ demonstrates that it is truly magnetic, being polarized in the $a$-$b$ plane, so it can be excited only by the $h$-component of the light. The absence of the electromagnon for $\vec{e} \parallel a, \vec{h} \parallel c$ also confirms its electric-dipole activity along the $c$-axis.

**TABLE I**. Energy $\hbar\omega$, oscillator strength $S_T$ (in the units of $\varepsilon_\infty$), and broadening $\gamma$ for the magnons (M1 and M2) and thee electromagnon (EM) modes measured at $T$=6 K in $h$-(Lu,Sc)FeO$_3$ with $d$ = 0.95 mm.

| $i$, $j$ | $\hbar\omega$, cm$^{-1}$ | $c$-axis $S_T = S_e$ | $\gamma$, cm$^{-1}$ | $\hbar\omega$, cm$^{-1}$ | $ab$-plane $S_T$ ($S_m$) | $\gamma$, cm$^{-1}$ |
|---|---|---|---|---|---|---|
| 1 | 22.5 (EM) | 0.009 | 1 | 26.5 (M1) | 0.01   (6·10$^{-4}$) | 5 |
| 2 | | | | 39 (M2) | 0.0005 (3·10$^{-5}$) | 3 |

**Figure 3(a)** shows circular-polarization resolved normalized transmission spectra for the light propagation along the $c$-axis. As in **Fig. 2(a)**, the σ+ and σ- spectra are dominated by the AFM resonance, or the M1 magnon with $k \cong 0$, at 0.8 THz. **Fig. 3(b)** displays the temperature dependence of the magnon peak which persists to about 140 K and thus spans the same temperature range as the canted ferromagnetic order. The second magnon M2 is too weak for systematic studies. In the following, we therefore focus mostly on the properties of the M1 magnon. A closer analysis of the transmission spectra revealed a doublet structure of this M1 resonance. The two components of the magnon, $\hbar\Omega^+(T)$ and $\hbar\Omega^-(T)$, which are active in the opposite circular polarizations of light $\sigma+$ and $\sigma-$, exhibit an unusually large splitting of about $\Delta$=0.06 THz. When measured with linearly polarized light, both components are simultaneously active and give rise to a broadened absorption peak in the transmission spectra. The temperature dependence of the magnon frequency [**Fig. 3(b)**] is also quite unusual for an antiferromagnetic material. Instead of a softening upon warming towards $T_N$, we observed an increase of the magnon frequency from 0.8 THz up to 1.0 THz when the temperature is increasing from 6 K approaching the AFM phase transition at $T$=160 K. Although the temperature dependence of the magnon energy resembles that for the magnetic susceptibility $\chi(T)$ from Ref. [3], there is no direct proportionality between them. Within the accuracy of our measurements, we did not detect any significant change in the splitting of the magnon doublet $\Delta = \hbar\Omega^+(T) - \hbar\Omega^-(T)$ with temperature [**Fig. 3(b)**]. The thermal cycling between 6 K and 300 K preserves the relative circular polarizations of the doublet components keeping $\hbar\Omega^+(T)$ above $\hbar\Omega^-(T)$, which indicates that the direction of the remnant magnetization $M_R$ that is responsible for the splitting is pinned to the sample even if the temperature is cycled up well above $T_N$ and back town to $T$=6 K. The dominant direction of the electric polarization $P$ of the sample is the main factor that determines the preferable remnant magnetization $M_R$ direction upon cooling. Note, however, that cooling in the external magnetic field of about $B$ =+1 T or $B$ =−1 T helps to reverse the preferred direction of $M_R$.

**Figure 4(a)** reveals that an external magnetic field $B$ further enhances the splitting of the M1 magnon. It shows that in the high-field regime (between 3 and 7 Tesla) the energy difference between two branches $\hbar\Omega^+(B,T)$ and $\hbar\Omega^-(B,T)$ is linear, with $\hbar\Omega^+(B,T) - \hbar\Omega^+(B,T) \approx g_{Fe3+}\mu_B B$, where $\mu_B$ = 0.013996 THz/T is the Bohr magneton. A good fit for the magnon energy dependence on the magnetic field and temperature has been obtained using the following empirical formula

$$\hbar\Omega^{\pm}(B,T) = \hbar\Omega_0(T) \pm \frac{1}{2}\sqrt{\Delta^2 + (g_{Fe3+}\mu_B B)^2} \, , \qquad (1)$$

where $\hbar\Omega_0(T)$ is the average for two curves shown in **FIG. 3(b)**. The g-factor of the magnon splitting is about $g_{Fe3+} = 3.0 \pm 0.5$ being comparable with that for magnons in other AFM compounds with $Fe^{3+}$ ions.[21]

The reversal of the magnetic field direction changes the selection rules for the two doublet components, as shown in **Fig. 4(b,c)**, such that the upper/lower branches have opposite circular polarizations. Note, however, that the change of the selection rules takes place at a finite reorientation field $B_R$ with a magnitude of about 0.25 T. The sudden change of the selection rules at $\pm B_R$ suggests that this process involves the reversal of the ferromagnetic spin canting and the corresponding direction of the sample magnetization. The temperature dependence of the reorientation field $B_R(T)$ is displayed in **Fig. 4(d)**. The measured magnitude of the magnon g-factor allows for estimation of the average remnant magnetization in the crystal at zero external field: $M_R = \Delta / g_{Fe3+}\mu_B = 1.4$ T. Thus, a relatively small external field of 0.25 T reverses magnetization $M_R$ that is equivalent to 1.4 T at the $Fe^{3+}$ sites of the $A_2$ spin structure.

At temperatures below 10 K, the directional asymmetry of $B_R$ provide evidence for a nonreciprocal light propagation. At $T$=4 K we observed that in one direction $B_R$ amounts to –0.2 T while in the other direction it is +0.4 T, as shown in **Fig. 4(d)**. A rotation of the sample by 180° with respect to the propagation direction of the light (while keeping it at the same low temperature) results in a corresponding inversion of the $B_R$ values. At the same time, the dominant selection rules for the circular dichroism of the $\hbar\Omega^{+}(B,T)$ and $\hbar\Omega^{-}(B,T)$ magnon branches remain unchanged. Both, the observation of $B_R$ reorientation and nonreciprocity of the circularly-polarized light propagation, or circular dichroism, is consistent with the symmetry of a system with internal macroscopic electric polarization $P$ and magnetization $M$ for the case of $\{\vec{k} \parallel \vec{M}_r, \vec{P}\} \neq \{\vec{k} \parallel -\vec{M}_r, -\vec{P}\}$ .

In the Raman spectra measured at $T$=5 K from the same single crystal of LSFO, a relatively broad peak was observed at the same energy as the magnon M1 (see the corresponding Section on Raman experiments). The infrared- and Raman-active magnons have a similar temperature dependence. The measured energies of the Raman-active magnons are shown in **FIG. 3(b)** with star symbols. The magnon is visible in the Raman scattering geometry with crossed circular polarizers for excitation and scattering $a(\sigma+, \sigma-)\bar{a}$ . No magnon signal was detected for the k-vector of the laser light along the c-axis.

In addition to the aforementioned circular dichroism at the magnon energies, we studied the interaction of the $A_2$ magnons with a THz vector vortex beam (VVB). **Figures 5(a,b)** show normalized transmission spectra in the vicinity of the M1 magnon for the clockwise and anticlockwise vorticity of the electric field vector, which corresponds to an orbital angular momentum (OAM) of $l$=+1 and $l$=−1, respectively. They reveal a splitting of the M1 magnon mode at zero field that gets enhanced by a magnetic field that is applied along the beam direction. **Figure 5(c)** shows the evolution of this magnon splitting for positive and negative magnetic fields and for OAM of $l$=+1 and $l$=−1. It reveals that the splitting of the M1 magnon peak can be inverted by changing the sign of the magnetic field or, likewise, that of the OAM. Note that this is the same kind of asymmetry with respect to time reversal as that for the circularly polarized THz beams [see **Fig.**

**4(b)]**, i.e. the sign of the handedness for circular polarization and the VVB is the same for the corresponding branches of the split magnon doublet. In contrast to the conventional transmission spectra for circular polarization, the VVB spectra demonstrate a stronger asymmetry in the line shape and a larger average amplitude of the absorption peak. Still, there are no strong arguments to relate the magnon line shape to the specifics of the ferroelectric domains or magnetic spin structure in LSFO. It is more reasonable to assign the observed asymmetry in **FIG. 5(a,b)** to systematic errors due to imperfect alignment of the axicon optics. More experimental studies are needed to develop the VVB spectroscopy into a practical tool for studies of the spin structures in antiferromagnetic materials.

We noticed that the low-temperature magnitude of the reorientation field $B_R$ reported earlier for the same LSFO crystals using magnetization measurements is 0.1 T,[3] which is different from our values of about 0.25 T [**Fig. 4(c)**]. The difference may be explained as due to the sample shape and size effect, which is significant for the slabs of $7 \times 7 \times 0.95$ mm$^3$ used in our THz experiments. At the same time, in Ref. [3] the zero-field magnetization for LSFO was found to be only $M_R = 0.01 \mu_B$ per formula unit. This result was explained due to small canting of the Fe spins in the $A_2$ phase. But, as showed above, the magnon splitting of 0.06 THz (or 2 cm$^{-1}$) corresponds to the magnetic field acting on the magnon-forming spin structure with Fe$^{3+}$ ions having nominal $S$=3/2 to be at least 50 times stronger. For comparison, we can mention that in $h$-$RE$MnO$_3$, no measurable zero-field splitting of the AFM magnons was observed,[22] which is, of course, expected for the $B_2$ spin structure that does not have any zero field magnetization. The calculated spectra of the refractive index and the Faraday rotation angle for LSMO that are based on the fit of the transmission spectra and magnon parameters in **Table I** are shown in Supplemental materials. The off-resonance value of the Faraday angle is calculated to be 0.25° for zero field $B$ and the sample thickness of 0.95 mm. Note that the magnon inhomogeneous broadening reduces the Faraday angle significantly compared to what can be expected for magnons with more narrow AFM resonances.

From the formal point of view, the situation with the strong zero-field dichroism of the magnons in LSFO that is observed for both, circular polarization and vector vortex beams is conceptually similar to the strong MOKE of 0.02° in altermagnetic Mn$_3$Sn with vanishing zero-field magnetization of $M_R = 0.002 \mu_B$/Mn.[23] In addition to MOKE, Mn$_3$Sn also demonstrated unusual resonant Faraday rotation and magnetoresistance,[24] and anomalous Hall effect[25], and all those effects are much larger than that may be expected from the almost negligible remnant magnetization. Thus, we may suggest a similarity between LSFO and other altermagnetic materials and with Mn$_3$Sn in particular. Then the strong dichroic effect is most likely related to the Berry's phase[26] that originates from $S_1 \cdot \left( S_2 \times S_3 \right) \neq 0$ inducing a large effective magnetic field that acts on the Fe non-coplanar spins $S_1, S_2$ and $S_3$ canted off the $a$-$b$ plane [**Fig. 1 (b)**]. The large zero-filed magnon splitting in LSFO should be additionally clarified with elaborate spin-structure calculations that includes the Dzyaloshinskii-Moriya interaction and deviation of the spin and orbital moments of the Fe ions off the $a$-$b$ plane. The interaction of magnons with the low-frequency phonons that may experience spin and orbital splitting at $k \neq 0$ is probably significant as well. Thus, we can attribute the observed strong magnon dichroism to the altermagnetic phase of LSFO. Correspondingly, we can predict a large MOKE in LSFO below 160 K.

### B. Optical phonons studied with far-IR ellipsometry

**Figure 6**(a,b) shows low-temperature experimental spectra of the pseudo dielectric function $\langle \varepsilon(\omega) \rangle$ measured with infrared ellipsometry for two orientations of the LSFO sample for which the $c$-axis is either in the plane of incidence of the light or perpendicular to it. In the first case, the pseudo dielectric function $\langle \varepsilon(\omega) \rangle$ is dominated by the TO phonons of the $A_1$ symmetry polarized along the $c$-axis, while in the other

case the spectra are dominated by the $E_1$ TO modes polarized in the $a$-$b$ plane. To extract parameters of the phonon modes, the ellipsometry spectra were fitted using an anisotropic dielectric function with Lorentz oscillators. In this framework, the dielectric function is parameterized as follows:

$$\varepsilon_{ab} = \varepsilon_{\infty,ab} + \sum_{i=1}^{n} \frac{\omega_{i,ab}^2 S_{i,ab}}{\omega_{i,ab}^2 - \omega^2 - i\gamma_{i,ab}\omega}$$

$$\varepsilon_c = \varepsilon_{\infty,c} + \sum_{j=1}^{m} \frac{\omega_{j,c}^2 S_{j,c}}{\omega_{j,c}^2 - \omega^2 - i\gamma_{j,c}\omega}$$

(2)

where $\omega_{i(j),ab(c)}$, $S_{i(j),ab(c)}$ and $\gamma_{i(j),ab(c)}$ are the transverse optical (TO) frequency, oscillator strength and inhomogeneous broadening of the $i^{th}$($j^{th}$) phonon mode polarized along the $ab$ plane ($c$-axis), and $\varepsilon_{\infty,ab(c)}$ is the value of the dielectric function along the $ab$ plane ($c$-axis) at frequencies higher than that of the highest-energy phonon mode. The extracted parameters of the phonon modes are listed in **Table II**.

In hexagonal crystals with the polar $P6_3cm$ space group, we expected to observe $9A_1$, $14$ $E_1$, and $15$ $E_2$ optical phonons, with $A_1$ and $E_1$ being both IR and Raman active, and $E_2$ only active in Raman experiments. The experimental data show 17 IR active phonon modes, 10 of which are polarized along the $c$-axis and 7 – in the $ab$-plane [see **Table II** and spectra in **Fig. 6(a,b,c,d)**]. One of the $c$-axis polarized $A_1$ phonons appears as a doublet at 205 cm$^{-1}$ and 223 cm$^{-1}$. This splitting may be understood as the local phase separation of LuFeO$_3$ - ScFeO$_3$. Among the $E_1$ modes that are polarized in the $ab$-plane we observed only 7 out of the expected 14 $E_1$ modes. This may be due to structural disorder-induced broadening of the modes which makes it difficult to spectrally resolve them. The high-frequency spectral features around 680 cm$^{-1}$ and 780 cm$^{-1}$ that are marked as LO modes and correspond to the well-known energy losses, or Berreman modes, that appear in the measured spectra of the pseudo-dielectric function of anisotropic samples when for one of the orthogonal components $\langle \varepsilon_1(\omega) \rangle$ is equal to zero. The LO modes do not contribute to the total count of the aforementioned IR-active modes. The experimental spectra in **FIG.6** were obtained using a stationary Silicon retarder that allowed us to obtain three Mueller matrix components for the measured LSFO sample. The corresponding spectra for the so-called N (or m$_{12}$), C (or m$_{33}$), and S (or m$_{43}$) Mueller matrix spectra[27] are shown in the Supplemental materials. Those spectra do not contain any additional information compared to what is already shown in **Fig. 6**, but they allow for a more accurate fit of the phonon parameters, all of which are summarized in **Table II**.

The calculated frequencies and symmetries of the IR- and Raman-active modes at 5 K (well below the magnetic ordering) in $h$-(Lu,Sc)FeO$_3$ are shown in **Table II**. The numbers are obtained using a linear interpolation of the calculated frequencies for the parent ternary compounds of LuFeO$_3$ and ScFeO$_3$ from Refs. [12,28]. The number of observed phonons doesn't change in the temperature range between 10 K and 300 K excluding any possible strong structural changes around the spin-ordering transition at about $T_N$(Fe)=160 K. The corresponding spectra between 10 K and 300 K for $\langle \varepsilon_c(\omega) \rangle$ and $\langle \varepsilon_{ab}(\omega) \rangle$ are almost overlapping, as shown in the Supplemental materials. Only the $A_1$ mode at 264 cm$^{-1}$, which is very weak, has an unusually strong temperature dependence that might be related to spin-structure changes in the AFM phase.

## C. Phonons studied with Raman scattering

The spectra of the optical phonons and magnetic excitations in the same LSFO crystals have been also studied using Raman scattering. Several back-scattering configurations were utilized: $a(c,c)\overline{a}$ for the $A_1$ modes, $a(b,b)\overline{a}$ and $c(a,a)\overline{c}$ for simultaneous observation of the $A_1$ and $E_2$ modes, $c(a,b)\overline{c}$ for the $E_2$ modes, and $a(c,b)\overline{a}$, $a(b,c)\overline{a}$ for $E_1$ modes. The first and the last symbols correspond to the $k$-vector direction for exciting and scattered light, whereas the symbols in brackets describe the polarization direction of the exciting and scattered light. Note that the in-plane $a$- and $b$-axes of the hexagonal lattice could not be distinguished here. In these geometries the mode symmetries $A_1$, $E_1$ and $E_2$ should become accessible in the $P6_3cm$ crystal structure of LSFO crystals as shown in brackets above. **Figure 7** shows the experimental Raman spectra, which allowed us to identify the positions of several $A_1$, $E_1$ and $E_2$ phonon modes. There are a total of 22 Raman active modes observed for LSFO at 5 K, among which are $9A_1$ modes, 6 $E_1$ modes and 7 $E_2$ modes. The low-temperature frequencies and the symmetries of the optical phonons are summarized in **Table II**. As expected for polar structures with $P \parallel c$, such as LSFO, there is a reasonable agreement between the $A_1$ phonon frequencies, which are active in both IR- and Raman scattering experiments. Several Raman modes around 260 cm$^{-1}$ in **FIG. 7(a)** are rather broad, which is most likely due to the disorder induced by the Lu-Sc compositional fluctuations.

In analogy to the IR experiments, the majority of the Raman-active phonon modes does not change between 5 K and 300 K, i.e., above and below $T_N$(Fe)=160 K. While several of the Raman peaks are quite broad, we could confirm that they exhibit a rather conventional, Grüneisen-like, temperature dependence with a plateau at low temperatures and a very weak softening as the temperature increases to 300 K. This indicates that the spin-phonon interaction in LSFO is quite weak as compared to other multiferroic compounds.

The low-frequency phonon $\omega_0$ =115 cm$^{-1}$ of the $E_2$ symmetry demonstrated an unusual asymmetry of the line shape for the Stokes and anti-Stokes parts of the spectrum [**Fig. 7(c)**] that is absent for the $A_1$ phonons in **Fig. 7(a)**. This asymmetry of the $E_2$ modes can be assigned to a Fano effect that appears in the Raman spectra due to coherent interaction between the sharp and discrete spectrum of the optical phonons and a continuum of polar excitations, which has a significant density of states $\rho$ at the optical phonon frequency.[29] We fitted the experimental lineshape of the phonon at $\omega_0$ =115 cm$^{-1}$ with the following Fano-type function

$$I(\omega) = A \frac{[q + E(\omega)]^2}{1 + E(\omega)^2} , \qquad (3)$$

where $E(\omega) = 2(\omega - \omega_0) / \gamma$, $\gamma$ is the inhomogeneous broadening, and $q$ is the asymmetry parameter, which turns out to be negative in our experiments. Note that $q \rightarrow \pm\infty$ corresponds to the symmetric Lorentz shape and the mode intensity $A \cdot q^2$ does not diverge because $A$ and $q$ are not independent parameters in the model. In theory $|q| \sim 1/\rho$, where $\rho$ is the density of states for the continuous excitations at the phonon frequency $\omega_0$.[30,31] The fitting yields that the magnitude of the Fano asymmetry parameter $q = -4.6 \cdot$ which also grows with the temperature increase. This indicates a dominant role of the two-phonon density of states of the acoustic modes. The continuum of carriers, like in Ref. [32], which should not have a strong temperature dependence, can be excluded based on the aforementioned THz and far-IR ellipsometry data, which show absolutely no sign of free-carrier absorption in this spectral range. A dynamic polarization buckling away from the exact $c$-axis and the corresponding rapid variation of the electric polarization in the $ab$-plane is one of the possible mechanisms of the Fano effect. Thus, the acoustic phonons coupled with polarization buckling may act together resulting in this optical phonon asymmetry.

**TABLE II**. Frequencies and symmetries of the experimental IR- and Raman-active modes at $T$= 10 K in $h$-(Lu,Sc)FeO$_3$. All frequencies are in units of cm$^{-1}$. The value of the dielectric function along the $ab$ plane ($c$-axis) $\varepsilon_{\infty,ab(c)}$ is measured at frequencies well above the highest phonon mode. The modes marked M correspond to magnetic modes. Calculated values are based on Refs. [12,28].

| | $c$-axis (9 $A_1$) | | | (a,b)-plane (14 $E_1$ and 15 $E_2$) | | | |
| | Calculated | IR | Raman | Calculated | IR | Raman | Raman |
| $\varepsilon_{\infty,ab}=4.5$ $\varepsilon_{\infty,c}=4.8$ | | $A_1$ | $A_1$ | $E_1$ | $E_1$ | $E_1$ | $E_2$ |
|---|---|---|---|---|---|---|---|
| M | | 22.5 (EM) | | 26.5 (M1) | | | |
| | | | | 39 (M2) | | | |
| 1 | | 111 | 109 | 134 | | | 115 |
| 2 | 158 | 152 | 155 | 171 | | | |
| 3 | 226 | 205, 223 | 240 | 181 | | | |
| 4 | 288 | 264 | 262 | 242 | | | |
| 5 | | | | 275 | | | |
| 6 | 301 | 302 | 284 | 313 | 302 | | |
| 7 | | | 317 | 325 | 322 | 319 | |
| 8 | 391 | 393 | | 387 | 343 | 348 | 410 |
| 9 | 423 | | | 402 | 422 | 422 | 422 |
| 10 | 448 | 442 | 440 | 431 | 434 | 496 | 496 |
| 11 | | 475 | | 482 | 483 | | 501 |
| 12 | 558 | 555 | 526 | 515 | 529 | 521 | 521 |
| 13 | 639 | | | 562 | | | |
| 14 | | 685(LO) | 657 | 627 | 770 (LO) | 657 | 657 |

### D. Electronic transitions studied with Ellipsometry and DFT+eDMFT

The electronic transitions of $h$-(Lu,Sc)FeO$_3$ in the near infrared–visible spectral ranges were measured at room temperature using spectroscopic ellipsometry. **Fig. 8(a,b)** shows experimental data for the real and imaginary parts of the pseudo dielectric function $\langle\hat{\varepsilon}(\omega)\rangle$ obtained at several discrete values for the angle of incidence in the range of 60° to 75°. **Fig. 8(c)** shows a sketch of the two experimental configurations of the ellipsometry measurements. In **Fig. 8**(a) the measured pseudo dielectric function $\langle\hat{\varepsilon}(\omega)\rangle$ is dominated by the $c$-axis component $\varepsilon_c(\omega)$, whereas in **Fig. 8** (b) it is dominated by $\varepsilon_{ab}(\omega)$. The average difference between the real parts of $\varepsilon_c(\omega)$ and $\varepsilon_{ab}(\omega)$ amounts to about 5%, with a maximum value of only about 20% near the strongest electronic resonance at 4 eV. Therefore, we used a quasi-isotropic model for each of the two configurations to fit the experimental data with a set of Lorentz oscillators that are polarized either along the $c$-axis, or in the $ab$-plane. This strategy of neglecting possible anisotropy effects is supported by the weak dependence of the spectra of $\langle \hat{\varepsilon}(\omega) \rangle$ on the angle of incidence. It is also in line with the analysis of the optical phonons in the infrared regime, as shown in **Fig. 6**. The fit parameters for the energies $E$, oscillator strengths $S$, and inhomogeneous broadening $\gamma$ are summarized in **Table III**. Note that the sum of the oscillator strengths for each configuration is close to the experimental low-energy values of $\varepsilon_{c,ab}(\omega)$ measured at 0.75 eV. The strongest electronic transitions at 4 eV polarized along the $c$-axis and at 3.7 eV in the $ab$-plane response dominate the spectral response of $h$-(Lu,Sc)FeO$_3$. In addition, as detailed in **Table III**, we have added electronic transitions with center frequencies outside of the measured spectral range, at 6.5 eV and 15 eV ($i,j$=7 and 8), as to obtain a better description of the high-energy part of the experimental spectra. The highest energy transition at 15 eV plays here the role of a so-called Penn gap that accounts for the mutual contributions of all higher-energy electronic transitions that are well-above the measured energy range.

**TABLE III**. Energy $E$, oscillator strength $S$ (in units of $\varepsilon_\infty$), and broadening $\gamma$ of the electronic transitions in $h$-(Lu,Sc)FeO$_3$.

| $i$, $j$ | $c$-axis | | | $ab$-plane | | |
|---|---|---|---|---|---|---|
| | $E$, eV | $S$ | $\gamma$, eV | $E$, eV | $S$ | $\gamma$, eV |
| 1 | 1.85 | 0.15 | 2.2 | 1.3 | 0.03 | 0.5 |
| 2 | 2.48 | 0.01 | 0.3 | 2.2 | 0.10 | 0.5 |
| 3 | 2.97 | 0.18 | 1.0 | 2.67 | 0.01 | 0.2 |
| 4 | 3.93 | 1.14 | 1.2 | 3.66 | 0.66 | 1.4 |
| 5 | 5.34 | 0.52 | 2.1 | 4.44 | 0.28 | 1.4 |
| 6 | | | | 5.35 | 0.30 | 1.6 |
| 7* | 6.5 | 0.5 | 1.9 | 6.3 | 0.25 | 1.4 |
| 8 Penn gap | 15 | 2.5 | 1.4 | 15 | 3.0 | 5 |
| | | $\sum_{j=1}^{8} S_j = 5 \pm 0.5$ | | | $\sum_{i=1}^{8} S_i = 4.7 \pm 0.5$ | |

To gain a deeper understanding of the electronic transitions derived from experimental data, we performed first-principles calculations using density functional theory combined with embedded dynamical mean-field theory (DFT+eDMFT), as implemented in the eDMFT code.[33,34,35,36,37,38] **Figure 9(a,b)** shows the computed dielectric function at 300 K (blue curves) for the $c$-axis and $ab$-plane, respectively. As seen from the comparison with the experimental data in **Figures 8(a,b)**, there is acceptable agreement between the main features that are highlighted by the black and blue arrows.

The Lu and Sc ion disorder present in the experimental sample significantly complicates the theoretical calculations. Therefore, we made several approximations in computing the dielectric function. It has been demonstrated in prior works that the eDMFT code can quantitatively predict the internal atomic coordinates within a given unit cell.[39,40,41] Based on this, we used the experimental unit cell of $h$-(Lu,Sc)FeO$_3$ (see Ref.[42]) and relaxed the internal coordinates separately for the LuFeO$_3$ and ScFeO$_3$ compounds, thereby obtaining theoretical crystal structures. These structures were subsequently used to calculate the electronic and optical properties of $LuFeO_3$ and $ScFeO_3$. The optical properties of $h$-$(Lu,Sc)FeO_3$ were then approximated by a weighted average of the individual dielectric functions (green and red curves in **Fig. 9(a,b)**). This method enables us to disentangle the contributions of Lu and Sc to the overall dielectric response. For instance, the lowest-energy electronic transitions, as well as those up to approximately 3.5 eV, appear to arise primarily from $LuFeO_3$, while transitions above this energy are predominantly due to $ScFeO_3$.

To understand the microscopic origin of these transitions, we also computed the electronic spectral functions and atom-projected density of states (PDOS). These are shown in **Fig. 9(c,d)** for $LuFeO_3$, and in **Fig. 9(e,f)** for $ScFeO_3$. The transitions below 4 eV originate mainly from hybridized $Fe(3d)$–$O(2p)$ states, while those above 4 eV arise from hybridized $O(2p)$–$Sc(3d)$ and $O(2p)$–$Lu(5d)$ states. Notably, the low-energy transitions from the top of the valence band to the bottom of the conduction band are dominated by $Fe(3d)$–$O(2p)$ hybridized states and are associated with nearly flat bands in the spectral function.

To further investigate these low-energy transitions, we plotted the orbital-projected density of states and the corresponding hybridizations between $Fe(3d)$ states and the $O(2p)$, $Lu(5d)$, and $Sc(3d)$ states in **Fig.10(a-e)**. In the $FeO_5$ bipyramidal environment characteristic of $h$-$(Lu,Sc)FeO_3$, the crystal field splitting in an ideal geometry yields two doublets [e(1) and e(2)] and one singlet a(1) for the $Fe(3d)$ states. However, due to the local distortion caused by the coexistence of Lu and Sc ions, all degeneracies are lifted. We therefore label the $Fe(3d)$-derived states as $e(1)_1$, $e(1)_2$, $e(2)_1$, $e(2)_2$, and $a(1)$, as shown in **Fig. 10(a)** and **(d)**. All $Fe(3d)$ projected states exhibit an energy gap, consistent with the insulating nature of the material. A detailed inspection of the self-energies for the $d$ states in both compounds reveals that four of these gaps are opened by strong local correlations, evidenced by prominent poles in the self-energies. These are referred to as *Mott gaps*. The fifth gap appears due to weaker correlations arising from covalent effects, characterized by a weak or absent self-energy pole. These are termed *correlated band gaps*. This analytical approach, based on the structure of self-energies within the gap, has also been used in previous studies to examine site- and orbital-selectivity in transition metal oxides.[40,41,43]

**Figures 10(c)** and **(f)** show strong hybridization between the $Fe(3d)$ singlet state and the $O(2p)$ states in both compounds, placing the Mott state at the top of the valence band alongside the oxygen states. The orbitals dominating the bottom of the valence band depend on whether Lu or Sc is present. While Sc induces only a minor distortion keeping $e(1)_1 \approx e(1)_2$ and $e(2)_1 \approx e(2)_2$, the nearly degenerate Lu induces a significant distortion. Empirically, one would assume a $Fe^{3+}$ oxidation state ($3d^5$ configuration), but due to the inclusion of realistic covalent effects in our calculations, the Fe ions exhibit an average oxidation state of $Fe^{2.5+}$. This is supported by the atomic microstate probabilities: although $3d^5$ remains the most probable configuration, there is a significant contribution from the $3d^6$ configuration, leading to an average $3d^{5.5}$ configuration. These covalent effects not only modify the oxidation state but also weaken the Mott gaps and induce spin fluctuations, thereby reducing the local effective magnetic moment in the paramagnetic state. To support this interpretation, we calculated the effective magnetic moment from the local magnetic susceptibility in the paramagnetic phase. Using theoretical crystal structure we obtain computed effective moments smaller than that would correspond for the high spin for the Fe ions in both $LuFeO_3$ and $ScFeO_3$, which aligns well with the experimentally observed value of 4.84 $\mu_B$.[42] The values of the effective moments that we obtained from theory are closer to the atomic moment for a spin state of $S$=2 or smaller, rather than the high-spin state of $S$ = 5/2 expected for $Fe^{3+}$. Recent studies on other compounds have also shown that covalent effects can lead to reduced effective spin values compared to those expected from formal oxidation states.[44] The low-temperature experiments and the corresponding DFT+eDMFT modeling for the magnetic phases of LSFO will be reported elsewhere.

## CONCLUSIONS

With THz- and Raman spectroscopy we have observed a number of magnetic excitations in hexagonal $h$-$Lu_{0.6}Sc_{0.4}FeO_3$ single crystals with non-collinear spins. The magnon modes are at 0.85 THz (M1) and at 1.2 THz (M2). The M1 mode consists of a doublet that is already split by about 0.06 THz at zero external magnetic field due to the presence of a remnant spontaneous magnetization. The M1 magnon has an unusual temperature dependence that resembles the behavior of the previously reported magnetic susceptibility. The observed circular and vector vortex beam dichroism allows us to attribute the M1 magnon to the $A_2$ spin structure, where dichroism is allowed due to the broken $\boldsymbol{P}\cdot\boldsymbol{T}$ symmetry. The latter is typical for the strong altermagnetic phases. In an external magnetic field $B$ we observed the M1 mode splitting with the $g$-factor of $Fe^{3+}$ ions being equal to 3.0 for the field direction along the $c$-axis. As expected only for the $A_2$ spin structure, the non-reciprocal propagation of light along the $c$-axis was observed for $T<10$ K. It revealed itself in the asymmetry of the reorientation field $B_R$ with respect to the preferable electric polarization of the sample. In addition to the magnons, an electromagnon polarized along the $c$-axis was found at 0.68 THz (EM). Our experiments did not allow us to attribute this excitation unambiguously to $A_1$ or $A_2$ spins configurations. Additional experiments for the THz light propagation in the $a$-$b$ plane in external magnetic field so that $\left(\vec{k} \perp \vec{B}, \vec{k} \perp \vec{P}\right)$ will be necessary to observe the symmetry allowed nonreciprocal effects for the $ab$-plane ordered spins in both the $A_1$ and $A_2$ spin phases. The IR and Raman-active optical phonons and their symmetries were identified. The phonons did not show any strong temperature dependence. The number of the observed $E_1$ and $E_2$ modes is less than expected for $h$-$Lu_{0.6}Sc_{0.4}FeO_3$ single crystals that is most likely due to the inhomogeneous broadening of the optical spectra. Several phonon modes may overlap being unresolved because of the Lu-Sc composition fluctuation. The low-frequency $E_2$ phonon at 115 cm$^{-1}$ showed a strong Fano asymmetry driven by the polarization fluctuations that are enabled by the acoustic phonons. The electronic transitions were measured in the infrared-to-ultraviolet parts of the optical spectrum. The electronic contribution to the dielectric function matches well the dielectric function values measured right above the optical phonon frequencies. In the further investigation of the magnetic properties of $h$-$(Lu,Sc)FeO_3$, special attention should be paid to the actual effective magnetic moment used in low-energy Hamiltonians, as it reflects important covalency and correlation effects beyond simple ionic models.

## ACKNOWLEDGMENTS

The authors are grateful to V. Kiryukhin , Y. Pashkevich, and D. Nykypanchuk for interest and useful discussions. Work at the New Jersey Institute of Technology and Rutgers University was supported by the U.S. Department of Energy under Contract No. DEFG02-07ER46382. The NSF MPS-ASCEND Award #2316535 supported the Raman scattering experiments and data analysis by V.A.M. F.L. acknowledges support from the Swiss National Science Foundation (SNSF) through an Early Postdoc Mobility Fellowship #P2FRP2-199598. Work at the National Synchrotron Light Source II at Brookhaven National Laboratory was funded by the DOE DE-AC9806CH10886. Use of the 22-IR-1(FIS) beamline was supported by the NSF EAR−2223273 (Synchrotron Earth and Environmental Science, SEES) and Chicago/DOE Alliance Center (CDAC) under the DOE-NNS cooperative agreement DE-NA-0003975. This research used the Materials Synthesis and Characterization facility of the Center for Functional Nanomaterials (CFN), which is a U.S. Department of Energy Office of Science User Facility, at Brookhaven National Laboratory under Contract No. DE-SC0012704. The work at the University of Fribourg was funded by the Swiss National Science Foundation through Grant No 200021-214905. G.L.P.'s work was supported by a grant of the Romanian Ministry of Education and Research, CNCS - UEFISCDI, Project No. PN-III-P1-1.1-TE-2019-1767, within PNCDI III. Computing resources for the theoretical calculations were provided by STFC Scientific Computing Department's SCARF cluster. Preparation of the input files and data processing for the theoretical calculations were performed on the local cluster at Stefan cel Mare University of Suceava (USV), obtained through a grant of the Romanian Ministry of Education and Research, CNCS— UEFISCDI, project number PN-III-P1-1.1-TE-2019-1767, within PNCDI III.

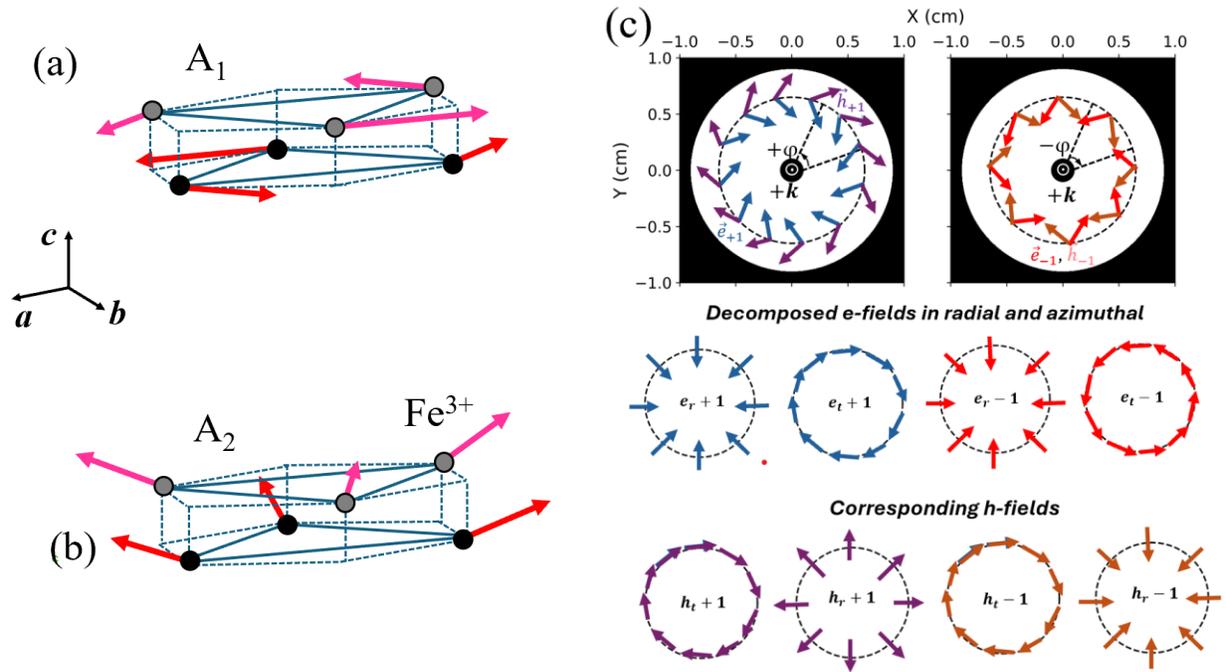

**FIG. 1** (a, b) The A1 and A2 spin structures for $Fe^{3+}$ magnetic ions in hexagonal LSFO (adopted from Refs. [1,10,11]. ). The $Fe^{3+}$ spins are co-planar in $A_1$ forming a toroidal spin structure. In $A_2$ the spins form a so called "magnetic monopole" and at the same time they produce a cantic ferromagnetic moment along the *c*-axis. Pink and red arrows correspond to the spins in two consecutive layers of $Fe^{3+}$ ions. Top of (c) shows the projections of electric $\vec{e}$ and magnetic $\vec{h}$ field distribution in the THz vector vortex beams produced with an axicon (adopted from Ref. [18]). The bottom panels present the coherent decomposition of the electric fields into the radial and azimuthal components. The bottom row shows the corresponding components of the *h*-fields that correspond to the e-fields in the middle row. Note that the AFM resonances or magnons are sensitive to the magnetic field part of these THz vector vortex beams.

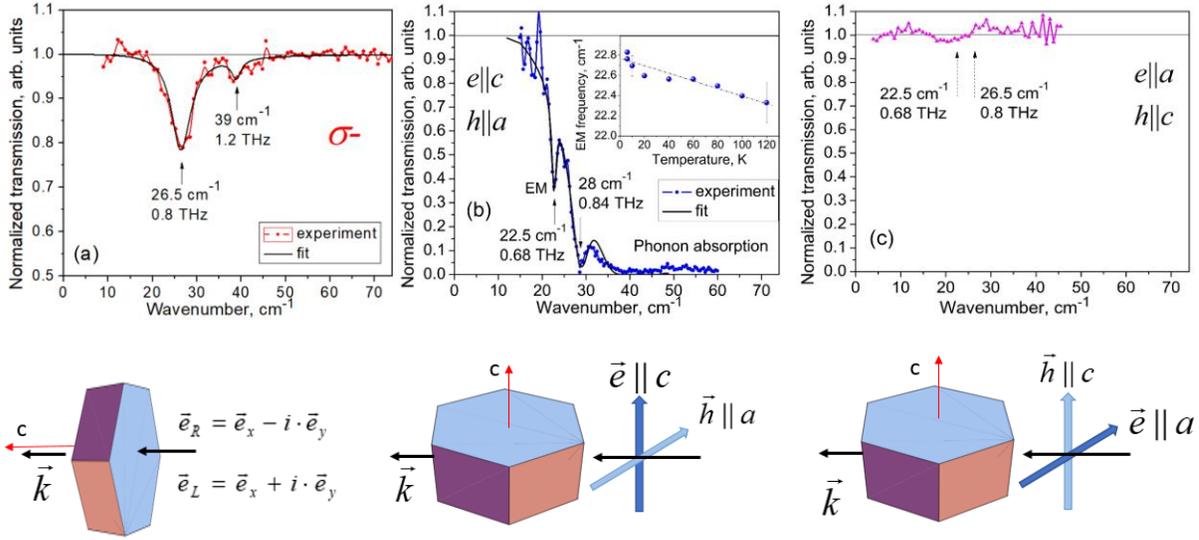

**FIG. 2** (a) Normalized transmission spectrum of LSFO at $B$=0 and $T$=6 K measured with anti-clockwise circular polarizations $\sigma-$ (red symbols). The light propagation is along the $c$-axis, the electric and magnetic fields of the light are in the $a$-$b$ plane. The solid line shows the best fit with two magnetic modes at 0.8 THz (26.5 cm⁻¹) and 1.2 THz (39 cm⁻¹) that are marked with arrows. The off-resonance transmittance value is 0.447, corresponding to $\varepsilon = 18$. (b) Normalized transmission spectrum at $B$=0 and $T$=6 K measured with linear polarizations $e \parallel c$, $h \parallel a$ (blue symbols). The light propagation is along the $b$-axis. The solid line shows the best fit with two magnetic modes at 22.5 cm⁻¹ and 28 cm⁻¹ that are indicated with arrows. The strong absorption at higher frequencies is due to the optical phonon at 115 cm⁻¹. The off-resonance transmittance value is 0.375, corresponding to $\varepsilon = 25$. (c) Normalized transmission spectrum of LSFO at $B$=0 and $T$=6 K with the linear polarizations $e \parallel a$, $h \parallel c$ shown with magenta symbols. The light propagation is along the $b$-axis. No magnetic modes were detected here. Vertical arrows at 22.5 cm⁻¹ and 26.5 cm⁻¹ mark the positions of the magnon modes in (a) and (b). Schematics for the measurement geometries are shown below each panel.

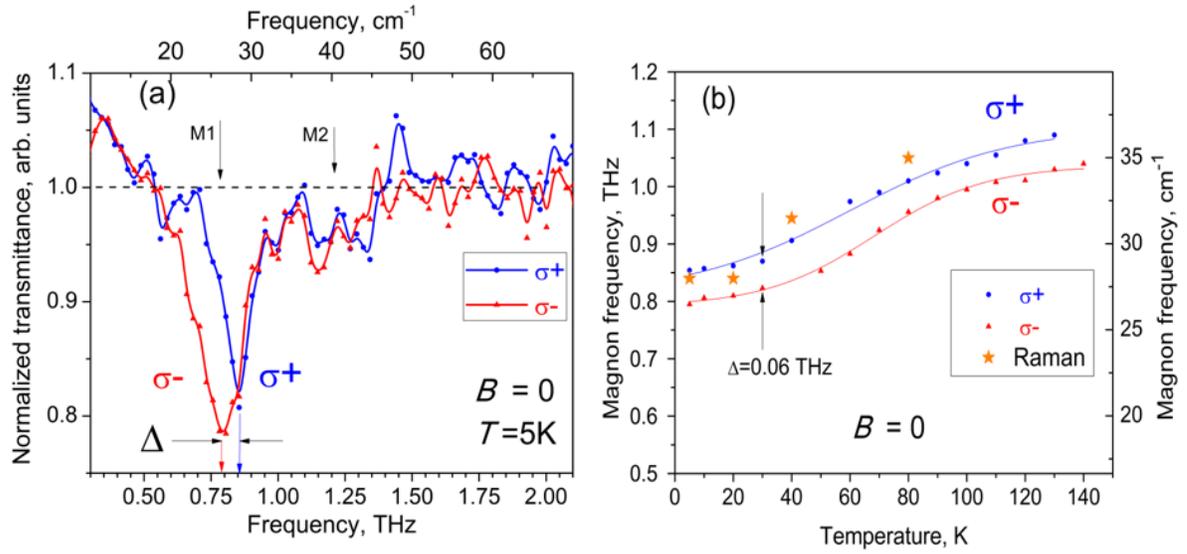

**FIG. 3** (a) The two normalized THz transmission spectra measured in LSFO at $B$=0 and $T$=6 K with two opposite circular polarizations, $\sigma+$ and $\sigma-$, shown with blue circles and red triangles, respectively. The mode splitting $\Delta$ is indicated with horizontal arrows. (b) Temperature dependence of the components of the magnon doublet M1 as measured with circularly polarized THz beams, $\sigma+$ and $\sigma-$, shown with blue circles and red triangles, respectively. Solid curves guide the eye. Orange stars show the corresponding magnon frequency in the Raman scattering data.

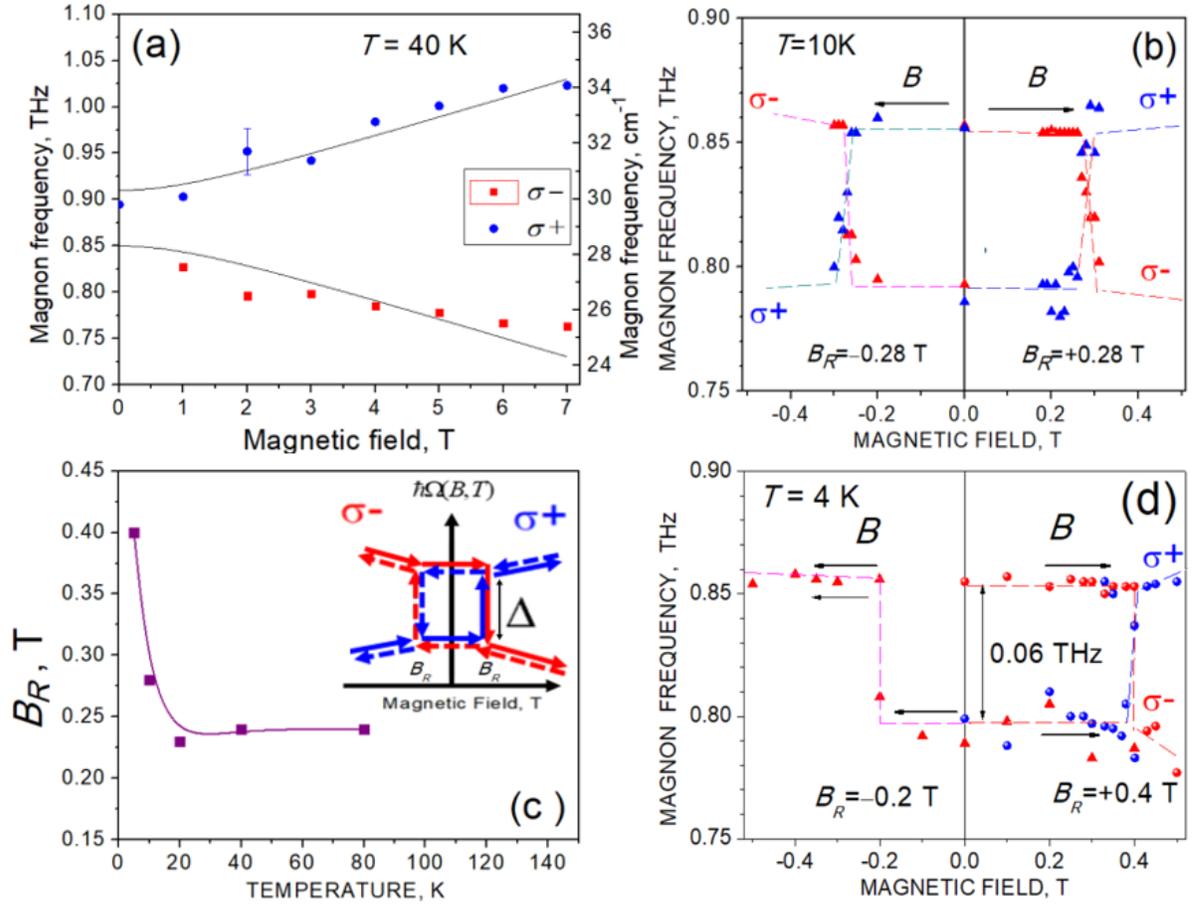

**FIG. 4** (a) Magnetic field dependence of the magnon doublet components measured with circularly polarized light $\sigma+$ and $\sigma-$ at $T$= 40 K. Solid curves are fits using Eq.(1) with $g_{Fe3+} = 3.0$ and a zero field splitting $\Delta = 0.06$ THz. The low field magnon frequencies measured with $\sigma+$ and $\sigma-$ are shown in (b) for $T$=10 K and in (d) for $T$=4 K. Dashed curves guide the eye. (c) Temperature dependence of the magnetization reversal field $B_R$. Schematics for the magnon selection rules in magnetic field is shown as an inset.

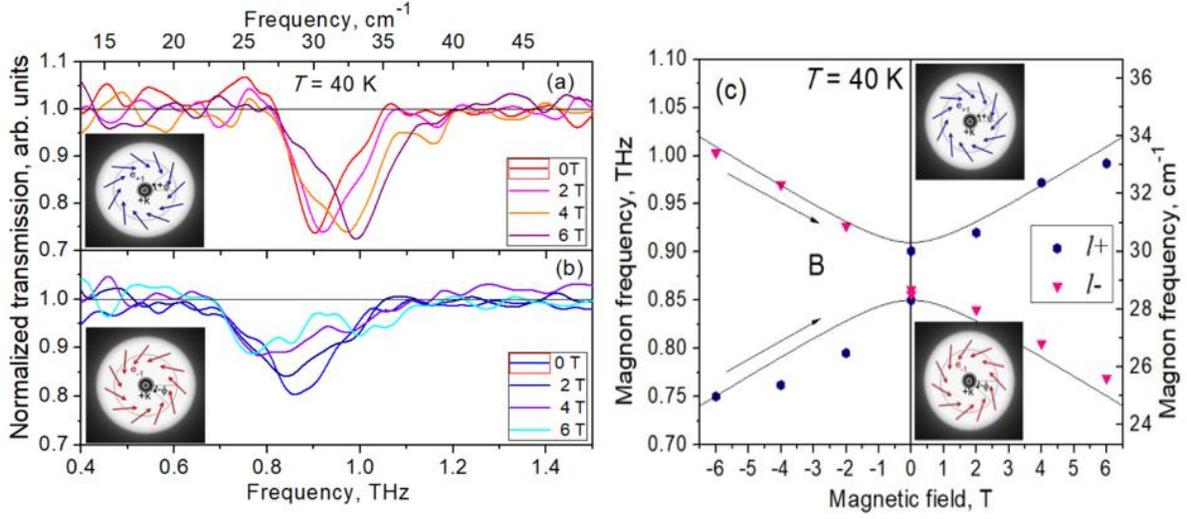

**FIG. 5.** Normalized THz transmission spectra of LSFO at positive B field and $T$=40 K measured with two opposite VVBs at (a) $l = +1$ and (b) $l = -1$. (c) Magnetic field dependence of the magnon doublet frequencies measured at $T$= 40 K with VVBs of $l+$ and $l-$ shown with magenta triangles and blue hexagons. Solid curves are fits using Eq. (1) with the same parameters as in **Fig. 4(a)**. The field is increased from $B$=−6T to $B$=+6 T during the measurement, as indicated with black arrows.

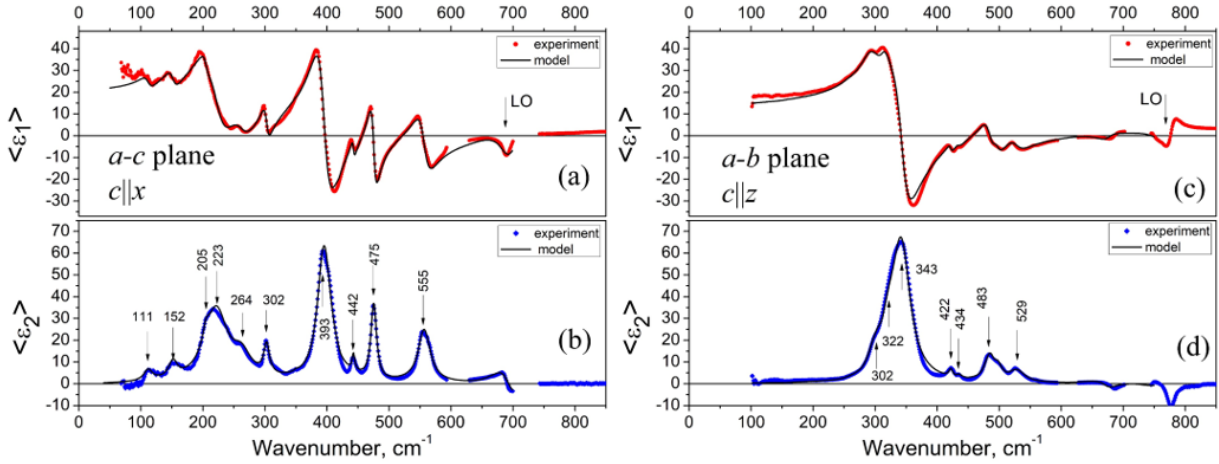

**FIG. 6.** Far-infrared ellipsometry spectra of the c-axis and ab-plane components of the anisotropic pseudo dielectric function $\langle \varepsilon_c(\omega) \rangle$ and $\langle \varepsilon_{ab}(\omega) \rangle$ of h-(Lu,Sc)FeO$_3$ measured at $T$=10 K with (a) the $c$-axis in the plane of reflection and (b) the $c$-axis perpendicular to the plane of reflection. The real parts $\langle \varepsilon_1(\omega) \rangle$ are shown with red and the imaginary parts $\langle \varepsilon_2(\omega) \rangle$ with blue symbols. The solid black curves are fits for an anisotropic dielectric function. The phonon peaks in (b) and (d) are marked according to their frequencies. The gaps in the experimental spectral range from 620 cm$^{-1}$ to 730 cm$^{-1}$ are due to light absorption in the silicon-made polarization optics, the beamsplitter, and polyethylene cryostat windows.

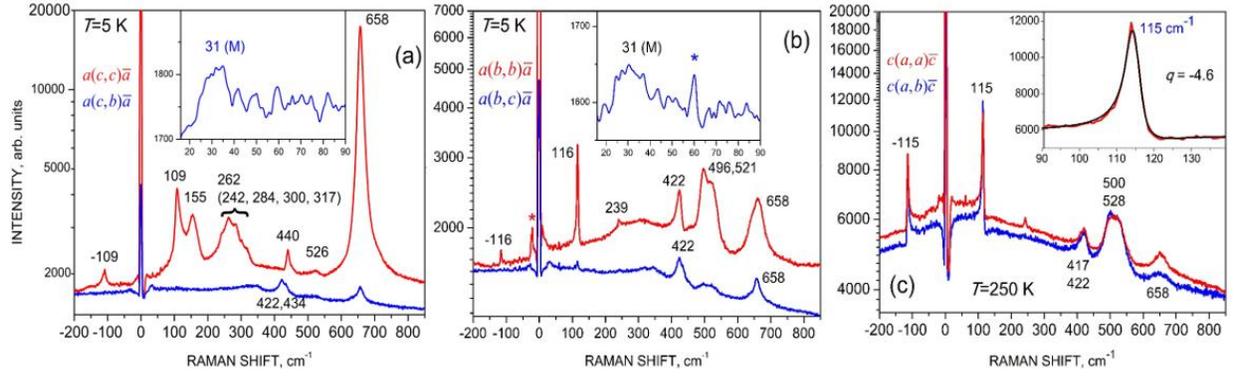

**FIG. 7.** Raman spectra of $h$-(Lu,Sc)FeO$_3$ measured in the complementary scattering geometries (a) $a(c,c)\overline{a}$ (red) and $a(c,b)\overline{a}$ (blue), (b) $a(b,b)\overline{a}$ (red), $a(b,c)\overline{a}$ (blue) and (c) $c(a,a)\overline{c}$ (red) and $c(a,b)\overline{c}$ (blue). Temperature is $T$=5 K in (a) and (b) and T=250K in (c). The phonon peaks are marked according to their frequencies. The insets zoom on the low-frequency parts of the corresponding Raman spectra with the magnon (M) peak at 31 cm$^{-1}$. Asterisks in (b) mark artifacts in the measured spectra due to insufficient straight light rejection. The Fano shape of the optical phonon at 115 cm$^{-1}$ in the inset of (c) was determined with an asymmetric Lorentzian fit (black curve) for an asymmetry parameter of $q$=−4.6.

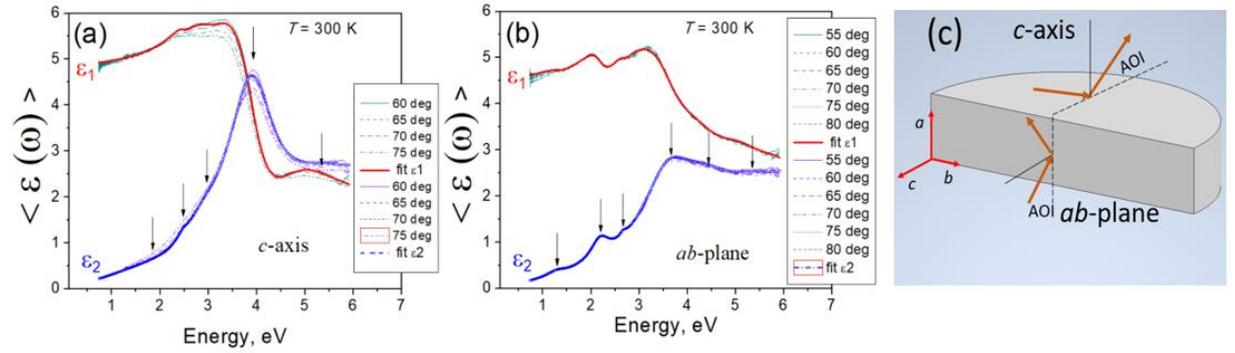

**FIG. 8.** Ellipsometry spectra of the pseudo-dielectric function $\langle\varepsilon(\omega)\rangle$ of $h$-(Lu,Sc)FeO$_3$ measured at $T$=300 K in the near-infrared to ultraviolet range. (a) Data for the $c$-axis in the reflection plane. (b) Data from the $ab$ plane with the c-axis being perpendicular to the sample surface The experimental spectra (cyan and violet curves) were taken at different AOIs between 55° and 80°. The fit using the Lorentz oscillator model are shown with red and blue spectra. The real $\varepsilon_1(\omega)$ and imaginary $\varepsilon_2(\omega)$ spectra are dominated by the electronic transitions whose energies are indicated by arrows. The experimental geometry for two measurements is shown in (c).

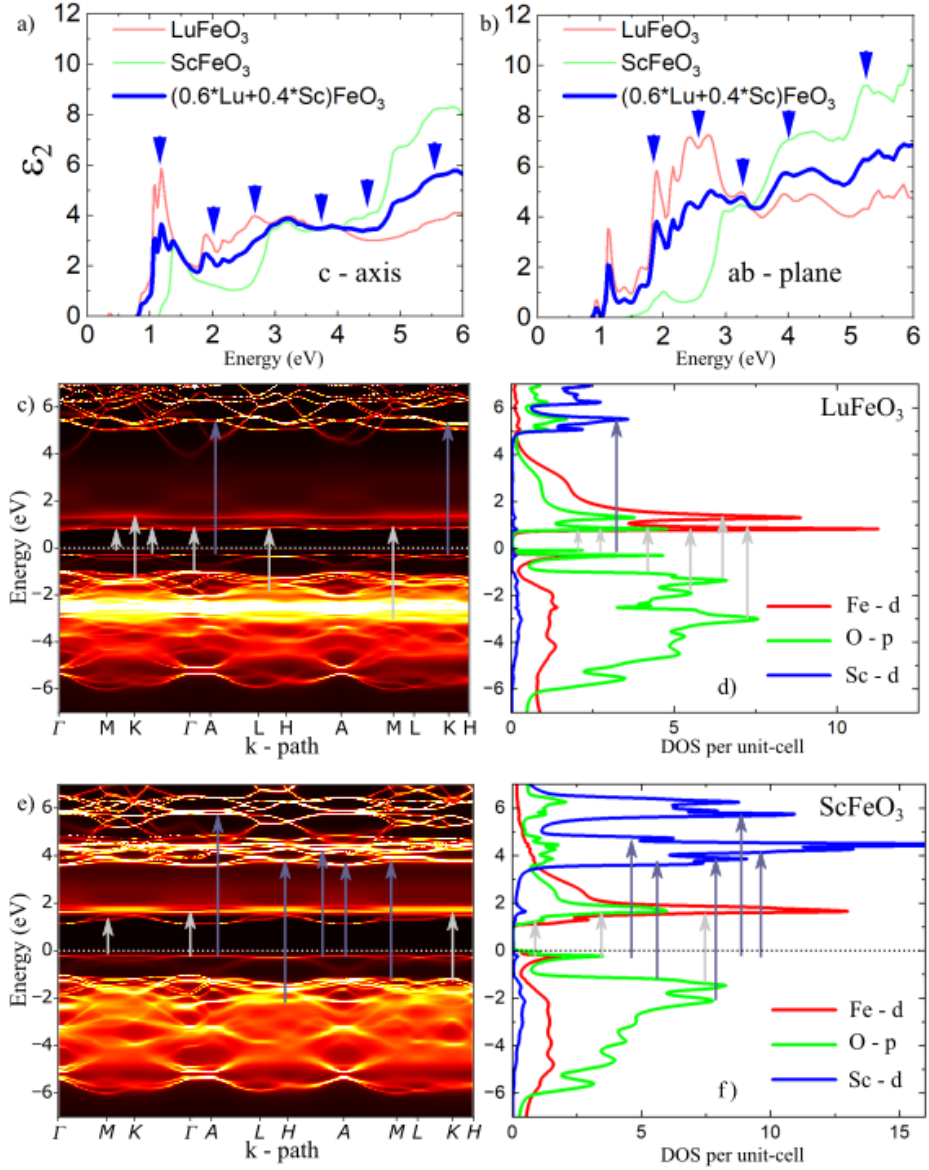

**FIG. 9.** Electronic properties of h-(Lu,Sc)FeO₃ computed using the DFT+eDMFT method at *T*=300 K. (a) and (b) Dielectric functions computed for the individual LuFeO₃ and ScFeO₃ compounds, based on theoretical structures obtained by relaxing the internal coordinates while keeping the experimental unit cell of h-(Lu,Sc)FeO₃ fixed. (c) and (e) Spectral functions along high-symmetry paths. (d) and (f) Orbital-projected density of states for Fe, O, Sc, and Lu atoms. The arrows in panels (a) and (b) indicate the positions of features observed in the experimental data (see Figure 8). Arrows in panels (c)–(f) schematically highlight potential electronic transitions identified in the optical measurements.

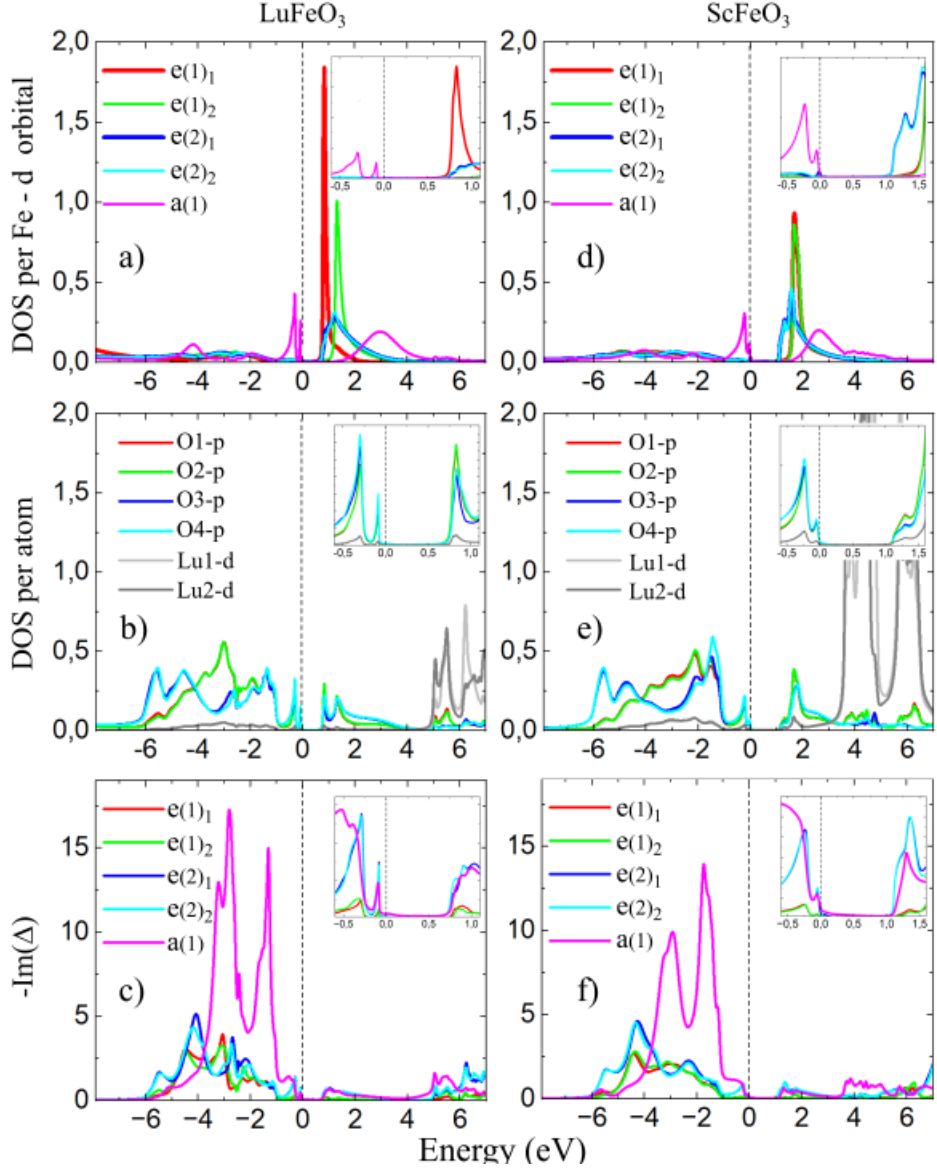

**FIG. 10.** Electronic properties of the Fe(3$d$), O(2$p$), Sc(3$d$), and Lu(5$d$) orbitals in $h$-(Lu,Sc)FeO₃, computed using the DFT+eDMFT method at 300 K. Panels (a), (b), and (c) show, respectively, the orbital-projected density of states (DOS) for Fe(3$d$), the atom-projected DOS for O(2$p$) and Lu(5$d$), and the imaginary part of the hybridization between Fe(3$d$) and O(2$p$)/Lu(5$d$) states for the theoretical LuFeO₃ compound. Panels (d), (e), and (f) display the corresponding quantities for the theoretical ScFeO₃ compound. Insets in all panels highlight the energy region around the band gap.